\documentclass[journal=jctcce, manuscript=article]{achemso}

\usepackage{chemformula} 
\usepackage[T1]{fontenc} 

\usepackage{physics}

\AbstractOn

\author{Lijie Ding}
\affiliation{Neutron Scattering Division, Oak Ridge National Laboratory, Oak Ridge, TN 37831, USA}
\author{Chi-Huan Tung}
\affiliation{Neutron Scattering Division, Oak Ridge National Laboratory, Oak Ridge, TN 37831, USA}
\author{Bobby G. Sumpter}
\affiliation{Center for Nanophase Materials Sciences, Oak Ridge National Laboratory, Oak Ridge, TN 37831, USA}
\author{Wei-Ren Chen}
\affiliation{Neutron Scattering Division, Oak Ridge National Laboratory, Oak Ridge, TN 37831, USA}
\email{chenw@ornl.gov}
\author{Changwoo Do}
\affiliation{Neutron Scattering Division, Oak Ridge National Laboratory, Oak Ridge, TN 37831, USA}
\email{doc1@ornl.gov}

\title{Off-Lattice Markov Chain Monte Carlo Simulations of Mechanically Driven Polymers}

\begin{document}

\begin{abstract}
We develop off-lattice simulations of semiflexible polymer chains subjected to applied mechanical forces using Markov Chain Monte Carlo. Our approach models the polymer as a chain of fixed-length bonds, with configurations updated through adaptive non-local Monte Carlo moves. This proposed method enables precise calculation of a polymer’s response to a wide range of mechanical forces, which traditional on-lattice models cannot achieve. Our approach has shown excellent agreement with theoretical predictions of persistence length and end-to-end distance in quiescent states, as well as stretching distances under tension. Moreover, our model eliminates the orientational bias present in on-lattice models, which significantly impacts calculations such as the scattering function, a crucial technique for revealing polymer conformation.
\end{abstract}
\clearpage

\section{Introduction}
Monte Carlo (MC) simulations have transformed polymer physics by enabling detailed exploration of polymer behavior across various scales, particularly where analytical methods are insufficient. These simulations efficiently navigate complex configurational spaces, yielding insights into polymer conformations and thermodynamics in diverse systems, from dilute solutions to complex melts. Both on-lattice and off-lattice models have been developed for simulation of polymers. MC simulations\cite{binder1997monte,binder2008recent,baschnagel2004monte} based on lattice models simulations simplify complexity, reduce computational demands, and align well with mean-field theories. For example, self-avoiding walk simulations on a cubic lattice have provided valuable insights into the configurational response of linear polymers under mechanical forces\cite{hsu2012stretching}. However, the on-lattice model introduces orientational bias, limiting its accuracy in calculating polymer responses to external forces when comparing with theoretical study\cite{chu2001small, mcculloch2013polymer, willner1994structural,auroy1991structures}. Off-lattice models\cite{theodorou2002variable,dodd1993concerted} resolve this issue by allowing movement in all directions in three-dimensional space, but existing off-lattice models do not account for bending, making them unsuitable for studying polymer deformation under external forces, which is important for designing materials with tailored mechanical properties and understanding experiments where bending is significant such as during stretching\cite{wang2011salient,li2016stretching} and steady shear flow\cite{smith1999single, schroeder2005dynamics, murphy2020capillary}

Meanwhile, one of the widely used technique to explore polymer conformations at molecular level under deformation is small angle scattering. Recent advancements in sample environments have enabled the application of mechanical energy within flow cells that is comparable to the bending energy of semiflexible polymers in solution. Accurately inverting the parameters that characterize the non-equilibrium configurations of mechanically driven polymer systems from the measured scattering functions presents significant challenges in data analysis. For precise quantitative spectral analysis, it is essential to integrate both the applied mechanical energy and the intrinsic bending energy of the polymers into the scattering functions. This process often relies heavily on MC simulations as computational references. To navigate the mathematical complexities inherent in analytical modeling, one promising approach leverages machine learning\cite{chang2022machine}. By employing computer simulations to generate an extensive library of scattering spectra based on various mechanical input parameters, it becomes feasible to establish correlations between these parameters and the unique spectral features, thereby laying the groundwork for regression analysis\cite{tung2022small}. For this approach to succeed, we need accurate calculations of polymer conformations. Since on-lattice models produces biased conformation due to the nature of lattice arrangement and the existing off-lattice models don't consider bending of a polymer chain, we develop an off-lattice model, inspired by the on-lattice simulations, that incorporate configuration updates involving the change of bending and efficiently samples the configuration space using Markov Chain Monte Carlo (MCMC) simulations. 

In this paper, we introduce the off-lattice MCMC model for semiflexible polymers where both internal bending energy and external field applied on the polymer can be accounted for. We carry out simulation of our system in three different states: in the absence of external field, under a uniaxial stretching and under a steady shear. When there is no external field, we demonstrate that our model produces unbiased scattering function unlike the on-lattice model and is in good agreement with theory when calculating conformation properties. When the polymer is under external force, we show that the deformation of the polymer is captured by the scattering function and quantify anisotropic conformation of polymers driven by the external field.

\section{Model}
We model the polymer as a chain of $N$ connected bonds, each of fixed length $l_b$. The orientation of bond $i$ is represented by $\mathbf{t}_i \equiv (\mathbf{r}_{i+1} - \mathbf{r}_i) / l_b$, where $\mathbf{r}_i$ denotes the position of the joint connecting bonds $i-1$ and $i$. One end of the polymer is anchored at the origin, $\mathbf{r}_0 = (0,0,0)$. We examine semiflexible polymer chains under uniaxial stretching and steady shear. For uniaxial stretching, the energy is given by 
\begin{equation}
    \label{equ:stretching}
    E_\mathrm{stretch} = \sum_{i=0}^{N-2} \frac{\kappa}{2}\frac{(\mathbf{t}_{i+1} - \mathbf{t}_i)^2}{l_b} - f X, 
\end{equation}
where $\kappa$ is the bending modulus, $f$ is the stretching force applied in the $x$-direction, and $X = (\mathbf{r}_N - \mathbf{r}_0) \cdot \mathbf{x}$ is the $x$-component of the end-to-end vector of the chain. For steady shear:
\begin{equation}
    \label{equ:shear}
    E_\mathrm{shear} = \sum_{i=0}^{N-2} \frac{\kappa}{2}\frac{(\mathbf{t}_{i+1} - \mathbf{t}_i)^2}{l_b} - \sum_{i=0}^{N-1} \gamma z_i (l_b \mathbf{t}_i \cdot \mathbf{x}).
\end{equation}
where $\gamma$ is the shear ratio along the $z$-direction, $z_i = \mathbf{r}_i \cdot \mathbf{z}$ is the $z$-component of joint $i$, and $(\mathbf{t}_i \cdot \mathbf{x})$ is the $x$-component of the bond tangent $\mathbf{t}_i$. Implicitly, self-avoidance is also considered by introducing a hard sphere interaction between polymer joints, with a sphere radius $R=l_b/2$.

\begin{figure}[!h]
    \centering
    \includegraphics{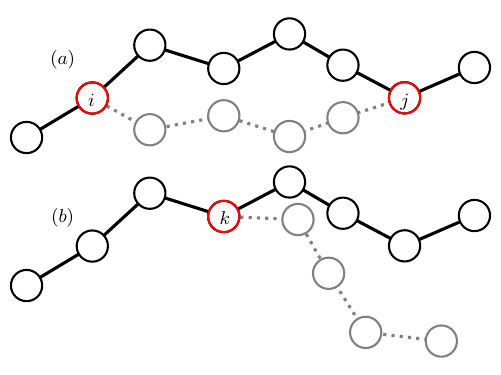}
    \caption{Monte Carlo updates of the polymer configuration: (a) Crankshaft around two joints, $i$ and $j$. (b) Pivot of the sub-chain from joint $k$ to the chain end.}
    \label{fig:config_update_demo}
\end{figure}

The innovation of our proposed approach lies in the techniques employed to generate and explore various polymer configurations through random sampling from the canonical distribution $\{\mathbf{r}_1, \dots, \mathbf{r}_N\} \sim \exp(-\beta E)$. Existing off-lattice Monte Carlo (MC) methods\cite{theodorou2002variable,dodd1993concerted} typically focus on polymers with fixed bond angles, which are unsuitable for mechanically driven polymer systems. Inspired by on-lattice MC motions, such as crankshaft and pivot moves \cite{verdier1962monte,kumar1988off,Binder1988}, we introduce two new configurational updates to address this issue.

For the off-lattice continuous crankshaft move, as shown in Fig.~\ref{fig:config_update_demo}(a), we select a sub-chain $(i, \dots, j)$ consisting of more than two bonds and rotate it around the axis defined by $(\mathbf{r}_j - \mathbf{r}_i)$ by a random angle within the interval $[-\phi_c, \phi_c]$. The crankshaft move keeps the chain's two ends fixed. To resolve the limitation of fixed ends, we propose an off-lattice pivot move that allows relative motion between the endpoints. In this move, as shown in Fig.~\ref{fig:config_update_demo}(b), a sub-chain starting from an endpoint $(k, \dots, N)$ is rotated around the bond at $k$ such that the tangent vector $\mathbf{t}_k$ is displaced to a random direction within the cone around the old $\mathbf{t}_k$, with an angle $\phi_p(k)$.

To improve the acceptance rate of the simulation, we use adaptive update variables. For the crankshaft move, we first randomly choose the contour length of the rotational part from the interval $[2, 2+\kappa]$, where the upper bound is proportional to the persistence length or bending modulus. We then randomly select the sub-chain $(i, \dots, j)$ with the chosen contour length. We also lower the rotation angles to support larger bending moduli or external fields, such that the crankshaft rotation angle is $\phi_c = \frac{2\pi}{3(1 + \kappa)}$. Additionally, the pivot rotation angle for joint $k$ increases as the joint approaches the end, such that $\phi_p(k) = \frac{\phi_c}{1 + (f + \gamma)(N - k)}$. These settings provide heuristic guidance for selecting the MC configuration updates. These updates are performed using the Metropolis-Hastings algorithm~\cite{metropolis1953equation,hastings1970monte}.

\section{Results}
We perform MC simulations of polymer chains with contour length $L = N l_b$. The simulation process is as follows: we initialize the chain along the $x$ direction and run the simulation for 1500 MC sweeps at inverse temperature $\beta \equiv 1 / k_B T = 0$ to allow for randomization. Subsequently, we increase $\beta$ gradually to 1 while running an additional 1500 MC sweeps. After this, we take measurements over another 3000 MC sweeps. Each MC sweep consists of $N^2$ crankshaft rotations and $N^2$ pivot rotations. In our simulations, we use natural units where lengths are measured in units of bond length $l_b$ and energies are measured in units of thermal noise $k_B T$. We firstly study the polymers in quiescent states where no external force is applied on the polymer, then turn to the polymers in stretching and shear.

\subsection{Polymers in the quiescent state}
In the quiescent states, the polymer configuration is solely determined by the contour length $L$ and bending modulus $\kappa$, and the scattering function should be spherically symmetric due to the lack orientational bias. The corresponding scattering function, expressed in terms of the two-point static correlation function in reciprocal space, is given by:
\begin{equation}
    I(\vb{Q}) = \frac{1}{N^2} \sum_{i=0}^{N-1}\sum_{j=0}^{N-1} e^{-i \vb{Q} \cdot (\vb{r}_i - \vb{r}_j)},
    \label{equ:2d_structure_factor}
\end{equation}
where $\vb{Q}$ is the scattering vector, and $N$ is the total number of segments. It is important to recognize that $I(\vb{Q})$ is inherently a three-dimensional quantity.
\begin{figure}[!h]
    \centering
    \includegraphics{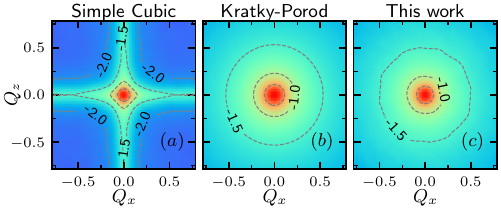}
    \caption{$I_{xz}(\vb{Q})$ for a semiflexible chain with $L = 200$ and $\kappa = 10$ in its quiescent state. The contour colors and levels represent $\log_{10}{I_{xz}(\vb{Q})}$. (a) Two-dimensional scattering intensity averaged over 3,000 polymer configurations generated on a simple cubic lattice using direct sampling. (b) Scattering intensity for the reference Kratky-Porod chain \cite{Landau}, calculated using numerical approximation\cite{pedersen1996scattering}. (c) Scattering intensity obtained using our proposed off-lattice approach.}
    \label{fig:2d_structure_factor}
\end{figure}

Experimentally, $I(\mathbf{Q})$ is not measured directly; rather, its projection onto a specific plane is collected. In Fig.~\ref{fig:2d_structure_factor}, we present the projection of $I(\vb{Q})$, calculated from different methods, onto the $Q_xQ_z$ plane for a semiflexible chain with contour length $L = 200$ and bending modulus $\kappa = 10$. This projection, defined as $I_{xz}(\vb{Q}) \equiv I(Q_x, Q_y = 0, Q_z)$, represents the intensity distribution from scattering experiment when the incident radiation is aligned along the $Q_y$ axis. The $I_{xz}(\vb{Q})$ derived from MC simulations using the simple cubic lattice model shows pronounced angular anisotropy as in Fig.~\ref{fig:2d_structure_factor}(a), which is a stark contrast to  the numerically calculated\cite{pedersen1996scattering} $I_{xz}(\vb{Q})$ of the reference self-avoiding Kratky-Porod chain\cite{kratky1949rontgenuntersuchung, Landau} shown in Fig.~\ref{fig:2d_structure_factor}(b). Given that a semiflexible chain, in the absence of external forces, should exhibit no preferred spatial orientation, this observed spectral anisotropy is clearly artificial. It originates from the spatial discretization inherent to the simple cubic lattice model, constrained by a coordination number of 6. In the context of MC simulations of mechanically driven polymer systems, such artificial scattering features could impair the accuracy of computationally generated static two-point correlation functions, which are essential for developing regression models to infer mechanical parameters from experimental spectra. On the contrary, as shown in Fig.~\ref{fig:2d_structure_factor}(c), our off-lattice approach circumvents the limitations of lattice models and produces an azimuthally symmetric $I(\vb{Q})$ that agrees well with the Kratky-Porod chain results in Fig.~\ref{fig:2d_structure_factor}(b). By eliminating spectral artifacts caused by spatial discretization, our method establishes a numerically accurate framework, facilitating the development of inversion algorithms for extracting mechanical properties of mechanically driven polymer chains from scattering spectra. This approach leads to a more precise depiction of polymer configurations and enables more accurate quantitative analysis of data obtained small-angle scattering experiments.

\begin{figure}[!th]
    \centering
    \includegraphics{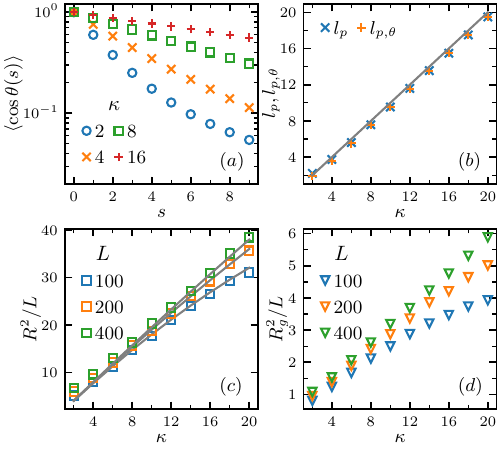}
    \caption{Effect of bending modulus $\kappa$ on polymer chains in quiescent states. (a) Bond angle correlation function $\left<\cos\theta(s)\right>$ as a function of $\kappa$ for a polymer length $L = 200$. (b) Persistence length $l_p$ versus $\kappa$ for $L = 200$, with the solid gray line indicating $l_p = \kappa$. (c) Normalized end-to-end distance of the polymer versus $\kappa$, with the solid gray line representing the theoretical prediction from Eq.~\eqref{equ:end-to-end-distance} where $t = e^{-1/\kappa}$. (d) Radius of gyration squared normalized by polymer length versus $\kappa$.}
    \label{fig:polymer_vs_kappa}
\end{figure}

Moreover, to assess the effectiveness of our method for calculating mechanical properties of the polymer, we study it's persistence length, end-to-end distance and radius of gyration. To quantify the bending effects, we compute the tangent-tangent correlation function $\left<\cos\theta(s)\right> = \left< \mathbf{t}_i \cdot \mathbf{t}_{i+s} \right>_i$, where the notation $\left< \cdots \right>_i$ denotes averaging over all bond pairs and s represents the contour distance between two bonds along a polymer chain. As shown in Fig.~\ref{fig:polymer_vs_kappa}(a), the correlation function $\left<\cos\theta(s)\right>$ initially exhibits an exponential decay. We model this decay to extract the persistence length $l_p$, utilizing the relation $\ln \left<\cos\theta(s)\right> = -s/l_p$. Furthermore, we determine the persistence length $l_{p,\theta}$ directly using the expression $l_{p,\theta} = -1/\ln \left<\cos\theta(1)\right>$. Fig.~\ref{fig:polymer_vs_kappa}(b) presents the calculated persistence lengths $l_p$ and $l_{p,\theta}$ (represented by colored symbols) as functions of the bending modulus $\kappa$, alongside the corresponding theoretical prediction~\cite{hsu2012stretching} shown as a solid gray line. The close quantitative agreement between our results and the theoretical model confirms that the persistence length scales linearly with the bending modulus, consistent with the relation $l_p = \kappa$. We further examine the squared end-to-end distance, $R^2$, which in the absence of self-avoiding effects is given by the theoretical expression~\cite{winkler1994models}:
\begin{equation}
    \label{equ:end-to-end-distance}
    \frac{R^2}{L} = \left[ \frac{1+t}{1-t} + \frac{2t}{L} \frac{t^L - 1}{(t - 1)^2} \right],
\end{equation}
where $t = \left<\cos\theta(1)\right>$, with natural units employed. Fig.~\ref{fig:polymer_vs_kappa}(c) illustrates the simulation results for $R^2$, which exhibit an increasing trend with the bending modulus $\kappa$, closely matching theoretical predictions for large values of $\kappa$ across various polymer lengths. Deviations at lower $\kappa$ are attributed to self-avoidance effects. Additionally, Fig.~\ref{fig:polymer_vs_kappa}(d) shows the scaled radius of gyration, $R_g^2/L$, where $R_g$ is the radius of gyration, which also increases with $\kappa$. It is important to note that $R^2/L$ cannot be estimated accurately through MC simulations based on lattice models. The quantitative agreement between our computed results and theoretical predictions, as displayed in Figs.~\ref{fig:polymer_vs_kappa}(b) and \ref{fig:polymer_vs_kappa}(c), underscores the robustness of our proposed approach.

\subsection{Polymers under external forces}
Having established and validated the framework of our approach for accurately predicting the configurations of semiflexible chains in their quiescent states, we now turn to evaluating its applicability for studying the configurations of semiflexible chains under mechanical deformations. Specifically, we investigate the behavior of the polymer under uniaxial stretching and steady shear.

\begin{figure}[!h]
    \centering
    \includegraphics{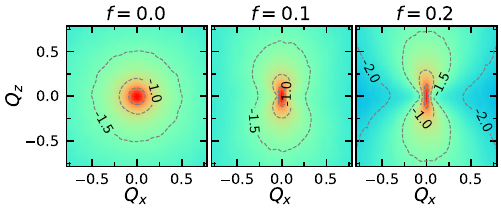}
    \caption{$I_{xz}(\vb{Q})$ for a semiflexible chain with $L = 200$, $\kappa = 10$ and $\gamma=0$ for stretching force $f=0, 0.1$ and $0.2$. The contour colors and levels represent $\log_{10}{I_{xz}(\vb{Q})}$.}
    \label{fig:Sq2D_f}
\end{figure}

When a stretching force $f$ is applied along the $x$ direction, the energy of the polymer is given by Eq.~\ref{equ:stretching}, where the extension of polymer's end from each other is favored. Thus the polymer's orientation symmetry is disrupted, which is reflected in the 2 dimensional scattering intensity as shown ins Fig.~\ref{fig:Sq2D_f}. When the stretching force $f=0$, the contour of $I_{xz}(\vb{Q})$ shows circular shape, indicating isotropy in the distribution of polymer configuration, as stretching force $f$ increases, the contour become oval-like and then dumbbell like orientate along the $Q_z$ direction, indicating the elongation of polymer along the $\mathbf{x}$ direction.

\begin{figure}[!h]
    \centering
    \includegraphics{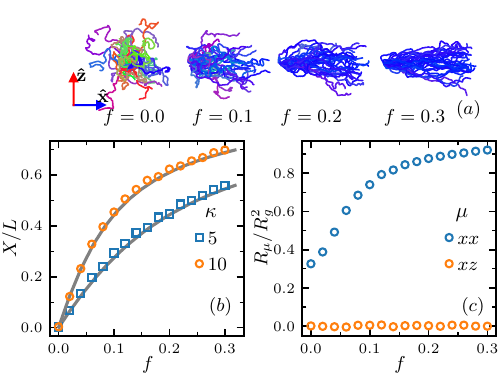}
    \caption{Polymer chains with $L=200$ subjected to uniaxial stretching force $f$. (a) Configurations of sampled polymers under varying $f$ with $\kappa=10$, color correspond to end-to-end orientation. (b) Normalized extension length along the $\mathbf{x}$-direction, $X/L$, versus stretching force $f$ for $\kappa=5$ and $10$. The solid gray line represents the theoretical curve based on Eq.~\eqref{equ: X vs ft}. (c) The $xx$ and $xz$ components of the gyration tensor, normalized by the square of the radius of gyration, plotted against the stretching force $f$.}
    \label{fig:polymer_vs_ft}
\end{figure}

The elongation of the polymers along the $x$-axis is also captured in Fig.~\ref{fig:polymer_vs_ft}(a), where the polymer extends progressively in the $x$-direction as $f$ increases. Theoretical approximation for this extension can be calculated using the asymptotic expression given by\cite{marko1995stretching}:
\begin{equation}
    \label{equ: X vs ft}
    f = \frac{1}{\kappa }\left[\frac{X}{L} + \frac{1}{4(1-X/L)^2} - \frac{1}{4}\right].
\end{equation}
As shown in Fig.~\ref{fig:polymer_vs_ft}(b), the extension length $X$ increases with the stretching force $f$ with the slope, spring constant increase with stretching force. And Our results show good agreement with the theoretical prediction provided by Eq.~\eqref{equ: X vs ft}. Additionally, we calculated the $xx$ and $xz$ components of the gyration tensor, where $R_{xx}=\frac{1}{2}\left<(x_i-x_j)^2\right>_{i,j}$ and $R_{xz}=\frac{1}{2}\left<(x_i-x_j)(z_i-z_j)\right>_{i,j}$, with $\left<\dots\right>_{i,j}$ denoting the average over all pairs of joints. When only the stretching force $f$ is applied, Fig.~\ref{fig:polymer_vs_ft}(c) shows that the ratio $R_{xx}/R_g^2$ increases from approximately $R_{xx}/R_g^2 \simeq 1/3$ at $f=0$ to nearly $R_{xx}/R_g^2 \sim 0.9$, while $R_{xz}/R_g^2$ remains flat at zero. This result aligns with the expectation that the polymer stretches exclusively along the $x$-direction, with no coupling between the $x$ and $z$ directions.

\begin{figure}[!h]
    \centering
    \includegraphics{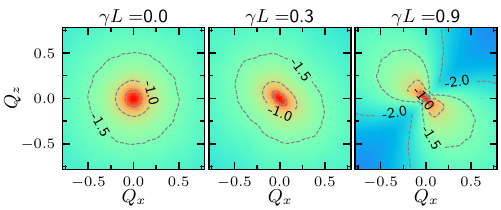}
    \caption{$I_{xz}(\vb{Q})$ for a semiflexible chain with $L = 200$, $\kappa = 10$ and $f=0$ for shear $\gamma L=0, 0.3$ and $0.9$. The contour colors and levels represent $\log_{10}{I_{xz}(\vb{Q})}$.}
    \label{fig:Sq2D_g}
\end{figure}

We further examine the behavior of the polymer chain under steady shear, where the polymer energy is given by Eq.~\ref{equ:shear}. As the shear is applied along the $\mathbf{z}$-direction with shear flow in the $\mathbf{x}$ direction, the shear force $\gamma$ induces deformation in the $xz$ plane, which is also captured by the scattering intensity as shown in Fig.~\ref{fig:Sq2D_g}. While in some degree similar to the case of stretching, the contour of $I_{xz}(\vb{Q})$ goes from circular shape to oval-like shape then dumbbell-like shape as the shear $\gamma$ increases, the $I_{xz}(\vb{Q})$ orientate along the $(Q_x,-Q_z)$ direction in stead of the $Q_z$ direction, indicating the deformation and elongation of the polymers is along the $(x,z)$ direction.

\begin{figure}[!th]
    \centering
    \includegraphics{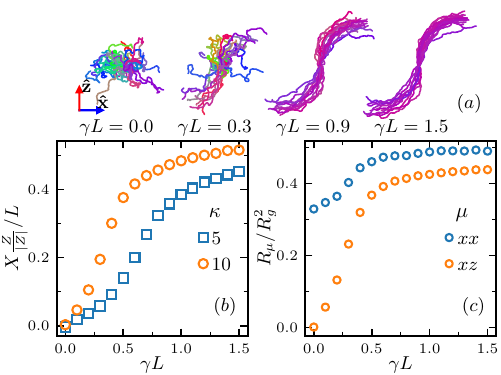}
    \caption{Polymer chains of $L=200$ and $\kappa=10$ subjected to shear $\gamma$. (a) Configurations of sampled polymers under varying shear $\gamma$, with $\kappa=10$ and no stretching force ($f=0$), color correspond to end-to-end orientation. (b) Normalized extension length of the polymer along the $\mathbf{x}$-direction versus shear $\gamma$. Due to the $\pm (X, Z)$ symmetry of the system, we use $XZ/|Z|$ for normalization. (c) The $xx$ and $xz$ components of the gyration tensor, normalized by the square of the radius of gyration, plotted against shear $\gamma$.}
    \label{fig:polymer_vs_fs}
\end{figure}

As illustrated in Fig.~\ref{fig:polymer_vs_fs}(a), the polymer extent along the $xz$ direction as the shear $\gamma$ increase, forming part of the S shape. We also calculate the polymer end-to-end extension. As shown in Fig.~\ref{fig:polymer_vs_fs}(b), the polymer extension $X/L$ shows a characteristic two-regime behavior: it increases rapidly with shear at low shear values, then more gradually as the shear $\gamma L$ approaches approximately 1.0. Additionally, the analysis of the gyration tensor reveals that the $xz$ component, $R_{xz}$, which was initially zero, increases with shear, indicating the symmetry breaking introduced by the shear force, as shown in Fig.~\ref{fig:polymer_vs_fs}(c).

\section{Summary}
In this paper, we present an off-lattice model for sampling the configuration space and calculating the conformational properties of semiflexible polymers under mechanical deformation using Markov Chain Monte Carlo simulations. Our model resolves the orientational bias rooted in on-lattice model and enables precise calculation of polymer's response to external force. Three states of the system have been studied, in the absence of external field, under an uniaxial stretching and in a steady shear. We demonstrate that the deformation of the polymer is captured by the scattering functions with expected anisotropic patterns and the polymer's conformation in response to external forces calculated using our model is in excellent agreement with theoretical prediction. 

Our model present a general framework for studying polymer conformation. While only bending and two cases of external field are discussed in this work, our model is applicable to any polymer energy as long as it can be written as a function of the polymer configuration. For instance, we can include torsion to study the chiral twisting of polymers, and when more complicated environments are created for the polymer, corresponding external forces can be modeled and applied, such as anisotropic stretching or non-uniform shear flows like Hagen-Poiseuille flow.  Moreover, our model can be extended to study polymer with more complicated shapes, i.e. star polymer and brush polymer or system with multiple polymers including polymer melt. For these system, due to the relative high probability of entanglement, we should further fine tune and lower the MC update range to improve sampling efficiency.

Additionally, this work paves way for future studies of polymer conformation through scattering function analysis. Recent advances in the Machine Learning-assisted scattering analysis\cite{chang2022machine} provide a general framework for inferring system parameters from scattering function using Machine learning (ML). Since our model generates unbiased scattering function, it can be used to create training sets for the ML algorithm and to analyze data from scattering experiments.

\begin{acknowledgement}
This research was performed at the Spallation Neutron Source and the Center for Nanophase Materials Sciences, which are DOE Office of Science User Facilities operated by Oak Ridge National Laboratory. This research was sponsored by the Laboratory Directed Research and Development Program of Oak Ridge National Laboratory, managed by UT-Battelle, LLC, for the U. S. Department of Energy. Monte Carlo simulations and computations used resources of the Oak Ridge Leadership Computing Facility, which is supported by the DOE Office of Science under Contract DE-AC05-00OR22725.
\end{acknowledgement}






\bibliography{achemso}
\end{document}